\documentclass[twocolumn,english]{revtex4-1}
\usepackage{ae,aecompl}
\usepackage[T1]{fontenc}
\usepackage[latin9]{inputenc}
\usepackage{geometry}
\usepackage{color}
\usepackage{babel}
\usepackage{float}
\usepackage{amsmath}
\usepackage{amssymb}
\usepackage{graphicx}
\usepackage[unicode=true,pdfusetitle,
 bookmarks=true,bookmarksnumbered=false,bookmarksopen=false,
 breaklinks=false,pdfborder={0 0 1},backref=false,colorlinks=true]
 {hyperref}
\begin{document}
\title{Second order divergence in the third order DC response of a cold semiconductor}
\author{G. B. Ventura}
\email{corresponding author: gbventura@fc.up.pt}

\author{D. J. Passos}
\affiliation{Centro de F\'isica das Universidades do Minho e Porto}
\affiliation{Departamento de F\'isica e Astronomia, Faculdade de Ci\^encias,
Universidade do Porto, 4169-007 Porto, Portugal}
\author{J. M. Viana Parente Lopes}
\affiliation{Centro de F\'isica das Universidades do Minho e Porto}
\affiliation{Departamento de F\'isica e Astronomia, Faculdade de Ci\^encias,
Universidade do Porto, 4169-007 Porto, Portugal}
\author{J. M. B. Lopes dos Santos}
\affiliation{Centro de F\'isica das Universidades do Minho e Porto}
\affiliation{Departamento de F\'isica e Astronomia, Faculdade de Ci\^encias,
Universidade do Porto, 4169-007 Porto, Portugal}
\begin{abstract}
In this work, we present the analytical expression for the second
order divergence in the third order DC response of a cold semiconductor,
which can be probed by different electric field setups. Results from
this expression were then compared, for the response of the gapped
graphene monolayer, with numerical results from a velocity gauge calculation
of the third order conductivity. The good agreement between the two
validates our analytical expression.
\end{abstract}
\maketitle

\section{Introduction}

The existence of divergences in the nonlinear optical (NLO) conductivities
of crystalline systems is a well-established result \citep{Aversa1995}.
It dates back to the discovery of the second order injection current,
whereby an ellipticaly polarized electric field generates a current
that, in the absence of saturation and relaxation, has a finite and
constant time derivative \citep{Aversa1996}. Shortly thereafter,
two other effects, the two-color current injection and the current
induced second harmonic generation, were described in terms of divergences
of the third order NLO conductivity \citep{Aversa1996,atasanov1996,vanDriel2001,khurgin1995}.
The first provided a valuable understanding of how the DC response
of crystals could be controled by dichromatic optical fields \citep{Aversa1996,atasanov1996,vanDriel2001},
while the second described the second harmonic generation in a material
where a DC field breaks the inversion symmetry of the crystal \citep{khurgin1995}.
The recent developments in two-dimensional materials \citep{Peres2009,novoselov2016,Carvalho2016,Manzeli2017}
and their optical properties \citep{mikhailov2007,mikhailov2008,Cheng2014,Mikhailov2016,Hipolito2016,Hipolito2017,ventura,passos,Ventura2019}
have spurred a new interest in these divergences, with works focusing
on effects such as the jerk current, the cross-phase modulation and
the degenerate four-wave mixing \citep{fregoso2018,fregoso2019,cheng2019}.
Such a systematic study of the divergences of the NLO conductivities
--- that depend on both a frequency sum and a relaxation rate ---
is important as they describe, in principle, responses that can be
expected to be \textit{large}, i.e., responses that should be easily
detectable and whose application can useful in the field of nonlinear
optics \citep{cheng2019}. 

As the second order DC response of a cold semiconductor carries a
divergence of first order in the inverse sum of frequencies and in
the phenomenological inverse relaxation rate, one can expect the third
order DC response to carry a divergence of second order --- in the
inverse of the sum of at least two of the frequencies, in the inverse
relaxation rate or in the product of both. These can be probed by
different setups of the electric field that have zero (or nearly zero)
sum of frequencies. The jerk current, recently proposed in ref.\citep{fregoso2018},
is the one associated to an electric field that has both a static
and monochromatic component; in this case, the divergence is in the
inverse square of the relaxation rate. Here, we shall also consider
two additional setups of the electric field through which the divergence
can be probed: one mixes the relaxation rate and a frequency sum and
is associated to the response to a \textit{dichromatic} field of frequencies
$\omega$ and $\delta\ll\omega$; the other involves the product of
two inverse frequency sums and is associated to the response to a
\textit{trichromatic} field of frequencies, $\omega+\delta_{1}\sim\omega$,
$\omega$ and $\delta_{2}\ll\omega$. The three different setups involve
output frequencies that are either zero, $\omega_{123}=0$, or small,
$\omega_{123}=\delta\ll\omega$, $\omega_{123}=\delta_{1}+\delta_{2}\ll\omega$,
and therefore fall under the scope of a DC (or quasi-DC) response.

The main point of this work, however, is that, regardless of the setup,
the divergence is always associated to the \textit{same} coefficient,
which is completely general and valid for a system with any number
of bands, whether in two or three dimensions. We note that the coefficient
derived here differs from the one derived in ref.\citep{fregoso2018}.
A second point regarding this expression is that, with exception of
the trichromatic setup, the pre-factor of the divergence depends on
the type of phenomenology used to include relaxation; it is not the
same when it is introduced via the equations of motion \citep{cheng2019}
or via adiabatic switching \citep{passos}. We use the latter one
in this work.

Finally, a word concerning the type of materials at hand: cold semiconductors.
Analytical calculations of NLO responses can be quite complicated,
as the number and diversity of different contributions increases quite
dramatically with the order of the calculation. These different contributions
are, in general terms, either dependent on the difference between
the occupation factors of two different bands, that is, a difference
of Fermi functions, or a derivative with respect to a Fermi function.
Cold semiconductors allow for a valuable simplification of this type
of calculation: all terms involving derivatives of Fermi functions
can be set to zero, while the occupation factors can be set to either
one --- in the valence bands --- or zero --- in the conduction
bands. 

This paper is organized as follows. In Sec.\ref{sec:II}, we derive
the terms of the conductivity that carry the second order divergence
in the DC (and quasi-DC) response. We also discuss how different electric
field setups probe this divergence in different ways, as well as the
role that phenomenology plays in the description of these divergent
responses. In Sec.\ref{sec:III}, we present a comparison between
the derived analytical results and results computed numerically in
the velocity gauge, for the gapped graphene monolayer \citep{passos,Ventura2019}.
The good agreement between the two validates the expressions that
we have derived. A brief summary of the work is presented in Sec.\ref{sec:IV}.

\section{The Third order response of a cold semiconductor and its second order
divergence in the dc response\label{sec:II}}

The nonlinear optical response of a crystalline system, which has
been the subject of extensive work \citep{Aversa1995,Aversa1996,vanDriel2001,atasanov1996,khurgin1995,Sipe2000,mikhailov2007,mikhailov2008,Cheng2014,Mikhailov2016,Hipolito2016,ventura,Hipolito2017,passos,joao2019,parker2019,Ventura2019,holder2019},
can be described --- at a given order in the electric field ---
in terms of certain response functions: susceptibilities, if the response
is expressed by the electric polarization; and conductivities, if
the response is expressed by the electric current. If we choose the
latter procedure to determine the response, and noting that the third
order response is the subject of interest in this work, one can write,
\begin{align}
J_{\beta}^{(3)}(t)= & \int\frac{d\omega_{1}}{2\pi}\frac{d\omega_{2}}{2\pi}\frac{d\omega_{3}}{2\pi}\sigma_{\beta\alpha_{1}\alpha_{2}\alpha_{3}}(\omega_{1},\omega_{2},\omega_{3})\nonumber \\
 & \times E^{\alpha_{1}}(\omega_{1})E^{\alpha_{2}}(\omega_{2})E^{\alpha_{3}}(\omega_{3})e^{-\imath\bar{\omega}_{123}t},
\end{align}
for the third order contribution to the electric current. The derivation
of $\sigma_{\beta\alpha_{1}\alpha_{2}\alpha_{3}}(\omega_{1},\omega_{2},\omega_{3})$
is a technical and laborious task that has been conducted in the aforementioned
works, for both the length \citep{Aversa1995,Aversa1996,atasanov1996,khurgin1995,Sipe2000,mikhailov2007,mikhailov2008,Cheng2014,Mikhailov2016,Hipolito2016,ventura,Hipolito2017},
and the velocity gauge \citep{passos,joao2019,parker2019,Ventura2019,holder2019}.
These gauge choices correspond to different ways of treating the coupling
between electrons in the crystal and the electric field, and follow
from the freedom that is involved in the choice of representation
of the electric field in terms of the scalar and vector potentials.
As we are interested in obtaining expressions that can be used in
analytical calculations, we choose to perform the calculations in
the length gauge. Moreover, we will carry over the notation that was
introduced in \citep{ventura}, as well as the notion that conductivities
can be fully determined by the energy bands, $\epsilon_{\mathbf{k}s}$,
and Berry connections, $\xi_{\mathbf{k}ss'}^{\alpha}$, of the electrons
in the crystal \citep{Aversa1995}.

The third order response is described by the following unsymmetrized
conductivity \citep{ventura},\begin{widetext}
\begin{align}
\frac{1}{ie^{4}}\sigma_{\beta\alpha_{1}\alpha_{2}\alpha_{3}}(\omega_{1},\omega_{2},\omega_{3})= & \int\frac{d^{d}\mathbf{k}}{(2\pi)^{d}}\sum_{s's}v_{\mathbf{k}s's}^{\beta}\frac{1}{\hbar\bar{\omega}_{123}-\Delta\epsilon_{\mathbf{k}ss'}}\bigl[D_{\mathbf{k}}^{\alpha_{3}},\frac{1}{\hbar\bar{\omega}_{12}-\Delta\epsilon_{\mathbf{k}}}\circ\bigl[D_{\mathbf{k}}^{\alpha_{2}},\frac{1}{\hbar\bar{\omega}_{1}-\Delta\epsilon_{\mathbf{k}}}\circ\bigl[D_{\mathbf{k}}^{\alpha_{1}},\rho_{\mathbf{k}}^{(0)}\bigr]\bigr]\bigr]_{ss'}.\label{eq:COND}
\end{align}
\end{widetext}Here, we consider that the $\bar{\omega}_{i}$ frequencies
contain a small imaginary part, $\bar{\omega}_{i}=\omega_{i}+i\gamma$,
for $\gamma$ the relaxation parameter. Note that frequencies with
multiple subscripts correspond to sums of frequencies: $\bar{\omega}_{12}=\bar{\omega}_{1}+\bar{\omega}_{2}$
and $\bar{\omega}_{123}=\bar{\omega}_{1}+\bar{\omega}_{2}+\bar{\omega}_{3}$.
As for $\Delta\epsilon_{\mathbf{k}ss'}$, it represents the energy
difference between two bands, $s$ and $s'$ at the same \textbf{k}-point
of the first Brillouin zone (FBZ), $\Delta\epsilon_{\mathbf{k}ss'}=\epsilon_{\mathbf{k}s}-\epsilon_{\mathbf{k}s'}$;
$\rho_{\mathbf{k}}^{(0)}$ is the density matrix in the absence of
a perturbation, $\rho_{\mathbf{k}ss'}^{(0)}=f_{\mathbf{k}s}\delta_{ss'}$.
We also note that $\circ$ represents the Hadamard, or element-wise,
product of two matrices in the band indexes, $(A\circ B)_{ss'}=A_{ss'}B_{ss'}$,
and that the integral is performed over the FBZ. The covariant derivative,
$D_{\mathbf{k}ss'}^{\alpha}$, and the velocity matrix elements, $v_{\mathbf{k}ss'}^{\alpha}$
are also defined as,
\begin{align}
D_{\mathbf{k}ss'}^{\alpha}= & \ \nabla_{\mathbf{k}}^{\alpha}\delta_{ss'}-i\xi_{\mathbf{k}ss'}^{\alpha},\label{eq:COV_DEV}\\
v_{\mathbf{k}ss'}^{\alpha}= & \ \frac{1}{\hbar}\bigl[D_{\mathbf{k}}^{\alpha},\mathcal{H}_{\mathbf{k}}\bigr]_{ss'},\\
= & \ \frac{1}{\hbar}\nabla_{\mathbf{k}}^{\alpha}\epsilon_{\mathbf{k}s}\delta_{ss'}-\frac{i}{\hbar}\Delta\epsilon_{\mathbf{k}s's}\xi_{\mathbf{k}ss'}^{\alpha},
\end{align}
for $\mathcal{H}_{\mathbf{k}ss'}=\epsilon_{\mathbf{k}s}\delta_{ss'}$.
The commutator of a covariant derivative with a matrix in band index
space of elements, $\mathcal{O}_{\mathbf{k}ss'}$, is given by the
expression,
\begin{align}
\bigl[D_{\mathbf{k}}^{\alpha},\mathcal{O}_{\mathbf{k}}\bigr]_{ss'}= & \bigl(\nabla_{\mathbf{k}}^{\alpha}\mathcal{O}_{\mathbf{k}ss'}\bigr)-i\bigl[\xi_{\mathbf{k}}^{\alpha},\mathcal{O}_{\mathbf{k}}\bigr]_{ss'}.\label{eq:COM_CD}
\end{align}
This means that for a cold semiconductor, where the Fermi-Dirac distribution
function reduces to $f_{\mathbf{k}v}=1$ in the valence bands and
$f_{\mathbf{k}c}=0$, in the conduction bands, the innermost commutator
in Eq.(\ref{eq:COND}) reads as,
\begin{align}
\bigl[D_{\mathbf{k}}^{\alpha_{1}},\rho_{\mathbf{k}}^{(0)}\bigr]_{ss'}= & -i\xi_{\mathbf{k}ss'}^{\alpha}\Delta f_{\mathbf{k}s's},
\end{align}
since $\bigl(\nabla_{\mathbf{k}}^{\alpha}f_{\mathbf{k}s}\bigr)=0$
for every band.

\subsection*{Second order divergence in the third order DC response}

We want to compute the second order divergence of Eq.(\ref{eq:COND})
in the case where the output frequency is either zero, $\omega_{123}=0$,
or is very small compared to the frequency of the optical component
of the field, $\omega_{123}\ll\omega$. To the second order divergence
of the third order DC conductivity we call $\Gamma_{\beta\alpha_{1}\alpha_{2}\alpha_{3}}$,
\begin{align}
\sigma_{\beta\alpha_{1}\alpha_{2}\alpha_{3}}(\omega_{1},\omega_{2},\omega_{3})= & \ \Gamma_{\beta\alpha_{1}\alpha_{2}\alpha_{3}}(\omega_{1},\omega_{2},\omega_{3})+(....),\label{eq:COND_EXP}
\end{align}
the ellipsis represents all other contributions to the conductivity.
To isolate this contribution, it is first useful to manipulate the
expression in Eq.(\ref{eq:COND}). We begin by exchanging the band
labels in the denominator that contains the total frequency, $\bar{\omega}_{123}$,
\begin{align}
\frac{1}{\hbar\bar{\omega}_{123}-\Delta\epsilon_{\mathbf{k}ss'}}\rightarrow & \ \frac{1}{\hbar\bar{\omega}_{123}+\Delta\epsilon_{\mathbf{k}s's}},
\end{align}
which allows us to express Eq.(\ref{eq:COND}) in the form 
\begin{multline}
\frac{1}{ie^{4}}\sigma_{\beta\alpha_{1}\alpha_{2}\alpha_{3}}(\omega_{1},\omega_{2},\omega_{3})=\\
=\int\frac{d^{d}\mathbf{k}}{(2\pi)^{d}}\mathsf{Tr}\bigl\{\bigl(v_{\mathbf{k}}^{\beta}\circ\frac{1}{\hbar\bar{\omega}_{123}+\Delta\epsilon_{\mathbf{k}}}\bigr)\left[D_{\mathbf{k}}^{\alpha_{3}},A_{\mathbf{k}}^{\alpha_{2}\alpha_{1}}\left(\bar{\omega}_{12},\bar{\omega}_{1}\right)\right]\bigr\}
\end{multline}
where the trace is taken over the band labels, and $A_{\mathbf{k}}^{\alpha_{2}\alpha_{1}}\left(\bar{\omega}_{12},\bar{\omega}_{1}\right)$
is the matrix in the commutator with $D_{\mathbf{k}}^{\alpha_{3}}$
in Eq.(\ref{eq:COND}). Upon using the cyclic invariance of the trace
($\mathsf{Tr}A[B,C]=-\mathsf{Tr}[B,A]C$) and integrating by parts
over $\mathbf{k}$, one can move the covariant derivative, $D_{\mathbf{k}}^{\alpha_{3}}$,
from acting on the terms on its right to acting on the terms on its
left,

\begin{widetext}
\begin{align}
\frac{1}{ie^{4}}\sigma_{\beta\alpha_{1}\alpha_{2}\alpha_{3}}(\omega_{1},\omega_{2},\omega_{3}) & =-\int\frac{d^{d}\mathbf{k}}{(2\pi)^{d}}\sum_{s's}\bigl[D_{\mathbf{k}}^{\alpha_{3}},v_{\mathbf{k}}^{\beta}\circ\frac{1}{\hbar\bar{\omega}_{123}+\Delta\epsilon_{\mathbf{k}}}\bigr]_{s's}A_{\mathbf{k}ss'}^{\alpha_{2}\alpha_{1}}\left(\bar{\omega}_{12},\bar{\omega}_{1}\right)\label{eq:LEFT}
\end{align}
with 
\[
A_{\mathbf{k}ss'}^{\alpha_{2}\alpha_{1}}\left(\bar{\omega}_{12},\bar{\omega}_{1}\right):=\frac{1}{\hbar\bar{\omega}_{12}-\Delta\epsilon_{\mathbf{k}ss'}}\bigl[D_{\mathbf{k}}^{\alpha_{2}},\frac{1}{\hbar\bar{\omega}_{1}-\Delta\epsilon_{\mathbf{k}}}\circ\bigl[D_{\mathbf{k}}^{\alpha_{1}},\rho_{\mathbf{k}}^{(0)}\bigr]\bigr]_{ss'}
\]
We proceed to separate the conductivity into its $s'\ne s$ and $s'=s$
contributions.

\end{widetext}

\subsubsection{$s'\protect\ne s$ contributions\label{subsec:Interband-contributions}}

For the $s'\ne s$ contributions, the integrand in Eq.(\ref{eq:LEFT})
reads as,
\begin{multline}
-\sum_{s'\ne s}\bigl[D_{\mathbf{k}}^{\alpha_{3}},v_{\mathbf{k}}^{\beta}\circ\frac{1}{\hbar\bar{\omega}_{123}+\Delta\epsilon_{\mathbf{k}}}\bigr]_{s's}\\
\times\frac{1}{\hbar\bar{\omega}_{12}-\Delta\epsilon_{\mathbf{k}ss'}}\bigl[D_{\mathbf{k}}^{\alpha_{2}},\frac{(-i)}{\hbar\bar{\omega}_{1}-\Delta\epsilon_{\mathbf{k}}}\circ\bigl[\xi_{\mathbf{k}}^{\alpha_{1}},\rho_{\mathbf{k}}^{(0)}\bigr]\bigr]_{ss'}.\label{eq:INTERBAND}
\end{multline}
We will show that these terms do not contribute to second order divergences
in the DC response. This requires expanding and manipulating the product
of commutators contained in Eq.(\ref{eq:INTERBAND}), as shown in
Appendix \ref{sec:B}. One term that follows from this procedure,
Eq.(\ref{eq:B4}), is, 
\begin{multline}
i\bigl[D_{\mathbf{k}}^{\alpha_{3}},v_{\mathbf{k}}^{\beta}\circ\frac{1}{\hbar\bar{\omega}_{123}+\Delta\epsilon_{\mathbf{k}}}\bigr]_{s's}\\
\times\xi_{\mathbf{k}ss'}^{\alpha_{1}}\Delta f_{\mathbf{k}s's}\bigl(\nabla_{\mathbf{k}}^{\alpha_{2}}\Delta\epsilon_{\mathbf{k}ss'})\frac{1}{\hbar\bar{\omega}_{12}-\Delta\epsilon_{\mathbf{k}ss'}}\frac{1}{(\hbar\bar{\omega}_{1}-\Delta\epsilon_{\mathbf{k}ss'})^{2}}.\label{eq:good_term}
\end{multline}
Consider the product of denominators that are associated to it,
\begin{equation}
\frac{1}{\hbar\bar{\omega}_{12}-\Delta\epsilon_{\mathbf{k}ss'}}\frac{1}{(\hbar\bar{\omega}_{1}-\Delta\epsilon_{\mathbf{k}ss'})^{2}}.\label{eq:DENOMINATOR}
\end{equation}
It has been suggested that terms of this form contribute to the second
order divergence in the third order DC response \citep{fregoso2018,fregoso2019}.
That, however, cannot be the case, since taking the limit of $\bar{\omega}_{2}\rightarrow0$
in Eq.(\ref{eq:DENOMINATOR}) shows us that the product of denominators
is not associated to any divergences,
\begin{align}
\frac{1}{\hbar\bar{\omega}_{12}-\Delta\epsilon_{\mathbf{k}ss'}}\frac{1}{(\hbar\bar{\omega}_{1}-\Delta\epsilon_{\mathbf{k}ss'})^{2}}\rightarrow & \ \frac{1}{(\hbar\bar{\omega}_{1}-\Delta\epsilon_{\mathbf{k}ss'})^{3}},\label{eq:no_mistake}\\
= & \ \frac{1}{2\hbar^{2}}\bigl(\partial_{\omega_{1}}^{2}\frac{1}{\hbar\bar{\omega}_{1}-\Delta\epsilon_{\mathbf{k}ss'}}\bigr),
\end{align}
and as such, the term in Eq.(\ref{eq:good_term}) cannot contribute
to $\Gamma_{\beta\alpha_{1}\alpha_{2}\alpha_{3}}$. A careful analysis
of all the other denominators contained in Eq.(\ref{eq:INTERBAND})
shows that no product of two divergent factors --- a second order
divergence --- appears for the DC (or quasi-DC) response. The divergent
factors in a conductivity with zero (or small) output frequency can
only come about when the energy difference, $\Delta\epsilon_{\mathbf{k}ss'}$,
in a denominator is zero, meaning that they only appear only when
the two band indexes are the same. We thus turn to the $s'=s$ contributions
of Eqs.(\ref{eq:LEFT}).

\subsubsection{$s'=s$ contributions\label{subsec:Intraband-contributions}}

For the $s'=s$ contributions, the integrand in Eq.(\ref{eq:LEFT})
reads as,
\begin{multline}
-\sum_{s}\bigl[D_{\mathbf{k}}^{\alpha_{3}},v_{\mathbf{k}}^{\beta}\circ\frac{1}{\hbar\bar{\omega}_{123}+\Delta\epsilon_{\mathbf{k}}}\bigr]_{ss}\\
\times\frac{1}{\hbar\bar{\omega}_{12}-\Delta\epsilon_{\mathbf{k}ss}}\bigl[D_{\mathbf{k}}^{\alpha_{2}},\frac{(-i)}{\hbar\bar{\omega}_{1}-\Delta\epsilon_{\mathbf{k}}}\circ\bigl[\xi_{\mathbf{k}}^{\alpha_{1}},f_{\mathbf{k}}\bigr]\bigr]_{ss}.\label{eq:INTRABAND}
\end{multline}
By expanding the two commutators, and after a careful treatment of
the terms involved --- presented in Appendix \ref{sec:A} --- one
can show that there is a single contribution involving the product
of two divergent factors in the case of a DC response,
\begin{equation}
-\frac{1}{\hbar\bar{\omega}_{123}}\frac{1}{\hbar\bar{\omega}_{12}}\frac{1}{\hbar}\sum_{r\ne s}\bigl(\nabla^{\beta}\nabla^{\alpha_{3}}\epsilon_{\mathbf{k}s}\bigr)\bigl[\frac{\xi_{\mathbf{k}sr}^{\alpha_{1}}\xi_{\mathbf{k}rs}^{\alpha_{2}}\Delta f_{\mathbf{k}rs}}{\hbar\bar{\omega}_{1}-\Delta\epsilon_{\mathbf{k}sr}}-(s\leftrightarrow r)\bigr],
\end{equation}
Here, $\Delta f_{\mathbf{k}rs}=f_{\mathbf{k}r}-f_{\mathbf{k}s}$.
By manipulating the band index sums and relabelling $r\rightarrow s'$,
we obtain
\begin{equation}
-\frac{1}{\hbar\bar{\omega}_{123}}\frac{1}{\hbar\bar{\omega}_{12}}\frac{1}{\hbar}\sum_{s\ne s'}\bigl(\nabla^{\beta}\nabla^{\alpha_{3}}\Delta\epsilon_{\mathbf{k}ss'}\bigr)\frac{\xi_{\mathbf{k}ss'}^{\alpha_{1}}\xi_{\mathbf{k}s's}^{\alpha_{2}}\Delta f_{\mathbf{k}s's}}{\hbar\bar{\omega}_{1}-\Delta\epsilon_{\mathbf{k}ss'}}.\label{eq:INTEGRAND}
\end{equation}
It is now clear how different frequency combinations of $\omega_{1}$,
$\omega_{2}$ and $\omega_{3}$, that is, different electric field
setups, correspond to different ways of probing the second order divergence
in the DC (or quasi-DC) response:\footnote{We note that this treatment of the divergences is similar to that
of a recent work, ref.\citep{cheng2019}, where the authors considered
the different divergences of the third order optical response of gapped
graphene.} 
\begin{itemize}
\item For $\omega_{123}=\omega_{12}=0$,\emph{ i.e.}, $\omega_{1}=-\omega_{2}$
and $\omega_{3}=0$,
\begin{align}
\frac{1}{\hbar\bar{\omega}_{123}}\frac{1}{\hbar\bar{\omega}_{12}}\rightarrow & \ \frac{1}{\hbar^{2}(3i\gamma)(2i\gamma)}=\frac{-1}{6\hbar^{2}}\gamma^{-2}.\label{eq:JERK_D}
\end{align}
This corresponds to the \textit{jerk} current: $\Gamma_{\beta\alpha_{1}\alpha_{2}\alpha_{3}}(\omega,-\omega,0)$.
\item For $\omega_{123}=\delta$, $\omega_{12}=0$, i.e., $\omega_{1}=-\omega_{2}=\omega$
and $\omega_{3}=\delta$, with $\gamma\ll\delta\ll\omega$,
\begin{align}
\frac{1}{\hbar\bar{\omega}_{123}}\frac{1}{\hbar\bar{\omega}_{12}}\rightarrow & \ \frac{1}{\hbar^{2}(\delta)(2i\gamma)}=\frac{-i}{2\hbar^{2}}\gamma^{-1}\delta^{-1}.\label{eq:MIXED_D}
\end{align}
This corresponds to the \textit{dichromatic} setup probe: $\Gamma_{\beta\alpha_{1}\alpha_{2}\alpha_{3}}(\omega,-\omega,\delta)$.
\item For $\omega_{123}=\delta_{1}+\delta_{2}$, $\omega_{12}=\delta_{2}$,
\emph{e.g.}, $\omega_{1}=\omega+\delta_{1}$, $\omega_{2}=-\omega$
and $\omega_{3}=\delta_{2}$, with $\gamma\ll\delta_{1},\delta_{2}\ll\omega$,
\begin{align}
\frac{1}{\hbar\bar{\omega}_{123}}\frac{1}{\hbar\bar{\omega}_{12}}\rightarrow & \ \frac{1}{\hbar^{2}(\delta_{1}+\delta_{2})(\delta_{2})}=\frac{1}{\hbar^{2}}(\delta_{1}+\delta_{2})^{-1}\delta_{2}^{-1}.\label{eq:DF_D}
\end{align}
This corresponds to the \textit{trichromatic} setup probe: $\Gamma_{\beta\alpha_{1}\alpha_{2}\alpha_{3}}(\omega+\delta_{1},-\omega,\delta_{2})$.
\end{itemize}
We can then express the denominator in its real and imaginary part
and, as $\hbar\gamma$ is the smallest energy scale in the integrand,
make the replacement,
\begin{figure}[t]
\includegraphics[scale=0.62]{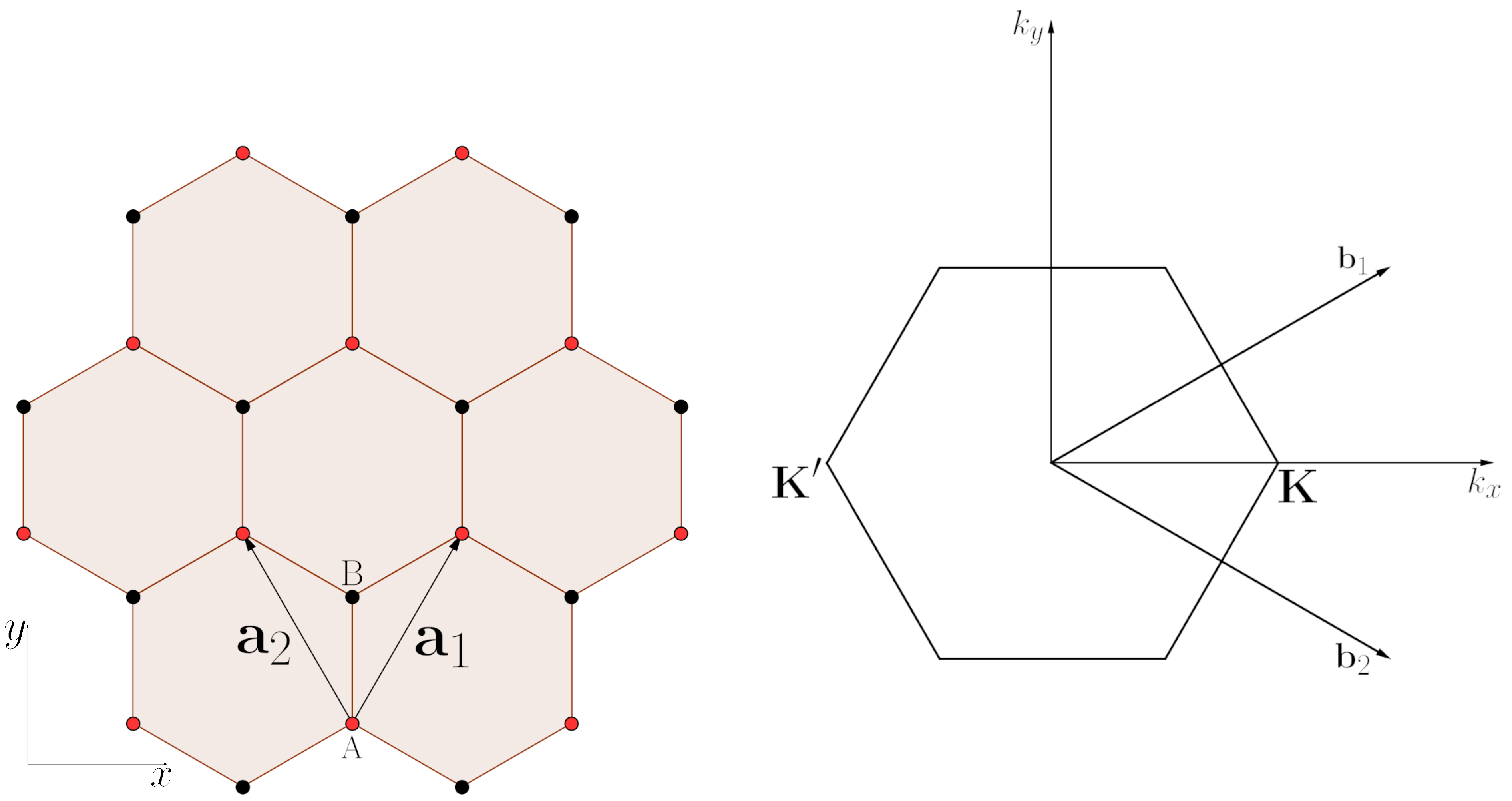}\caption{The honeycomb lattice of the gapped graphene monolayer and its associated
FBZ. Left: lattice structure of the gapped graphene monolayer. Note
that the A and B sites are not equivalent and have different on-site
energies, $\epsilon_{\text{A}}-\epsilon_{\text{B}}=\Delta$. Right:
the first Brillouin zone and the corresponding reciprocal lattice
vectors. The band minimums are located at the vertices of the Brillouin
zone which correspond to the $\mathbf{K}$ and $\mathbf{K}'=-\mathbf{K}$
points. \label{fig:1}}
\end{figure}
\begin{align}
\frac{1}{\hbar\bar{\omega}_{1}-\Delta\epsilon_{\mathbf{k}ss'}}\rightarrow & \ \frac{P}{\hbar\omega_{1}-\Delta\epsilon_{\mathbf{k}ss'}}-i\pi\delta(\hbar\omega_{1}-\Delta\epsilon_{\mathbf{k}ss'}).
\end{align}
and write the contribution to the conductivity that carries the second
order divergence, $\Gamma_{\beta\alpha_{1}\alpha_{2}\alpha_{3}}$,
as,
\begin{multline}
\frac{ie^{4}}{\hbar^{3}}\frac{(-1)}{\bar{\omega}_{123}\bar{\omega}_{12}}\int\frac{d^{d}\mathbf{k}}{(2\pi)^{d}}\sum_{s\ne s'}\bigl(\nabla^{\beta}\nabla^{\alpha_{3}}\Delta\epsilon_{\mathbf{k}ss'}\bigr)\xi_{\mathbf{k}ss'}^{\alpha_{1}}\xi_{\mathbf{k}s's}^{\alpha_{2}}\Delta f_{\mathbf{k}s's}\\
\times\bigl[\frac{P}{\hbar\omega_{1}-\Delta\epsilon_{\mathbf{k}ss'}}-i\pi\delta(\hbar\omega_{1}-\Delta\epsilon_{\mathbf{k}ss'})\bigr].\label{eq:UNISYM}
\end{multline}
Note that we have not yet symmetrized the conductivity. As we are
interested in the terms that contribute to the second order divergence
in the DC response, $\Gamma_{\beta\alpha_{1}\alpha_{2}\alpha_{3}}$,
there is a single relevant permutation, $(\alpha_{1},\omega_{1})\leftrightarrow(\alpha_{2},\omega_{2})$,
that is to be taken into account. After a careful calculation ---
see Appendix \ref{sec:C} --- one obtains a symmetrized $\Gamma$,
$\tilde{\Gamma}_{\beta\alpha_{1}\alpha_{2}\alpha_{3}}$, 
\begin{align}
\tilde{\Gamma}_{\beta\alpha_{1}\alpha_{2}\alpha_{3}}(\omega_{1},\omega_{2},\omega_{3})= & -\frac{e^{4}}{3\hbar^{3}}\frac{1}{\bar{\omega}_{123}}\frac{1}{\bar{\omega}_{12}}\iota_{\beta\alpha_{1}\alpha_{2}\alpha_{3}}(\omega_{1}),\label{eq:SECOND_ORDER_DIVER}
\end{align}
with a coefficient, $\iota_{\beta\alpha_{1}\alpha_{2}\alpha_{3}}(\omega_{1})$,
that is expressed in terms of a single integral involving a Dirac
delta function,\footnote{Note that, as we are specifically dealing with the second order divergence,
the $\iota$ coefficient will effectively be a function of the main
optical frequency, $\omega$.} 
\begin{multline}
\iota_{\beta\alpha_{1}\alpha_{2}\alpha_{3}}(\omega_{1})=\ \pi\int\frac{d^{d}\mathbf{k}}{(2\pi)^{d}}\sum_{s\ne s'}\bigl(\nabla^{\beta}\nabla^{\alpha_{3}}\Delta\epsilon_{\mathbf{k}ss'}\bigr)\\
\times\xi_{\mathbf{k}ss'}^{\alpha_{1}}\xi_{\mathbf{k}s's}^{\alpha_{2}}\Delta f_{\mathbf{k}s's}\,\delta(\hbar\omega_{1}-\Delta\epsilon_{\mathbf{k}ss'}).\label{eq:SYM}
\end{multline}
\\
Finally, we must comment on the numerical pre-factors in $\tilde{\Gamma}_{\beta\alpha_{1}\alpha_{2}\alpha_{3}}$,
Eq.(\ref{eq:SECOND_ORDER_DIVER}). For the jerk current and the dichromatic
setup probe to the divergence, different choices of phenomenology
are associated to different numerical factors. If, for example, we
were to consider the relaxation rate introduced via equations of motion,
one would have $\bar{\omega}_{1}=\ \omega_{1}+i\gamma$, $\bar{\omega}_{12}=\ \omega_{12}+i\gamma$,
and $\bar{\omega}_{123}=\ \omega_{123}+i\gamma$, and factors of $1/6$
and $1/2$ would not appear in Eq.(\ref{eq:JERK_D}) and in Eq.(\ref{eq:MIXED_D}),
respectively. We will retain the use of the adiabatic switching approach,
as we want to compare analytical results with results from a numerical
calculation of the conductivity in the velocity gauge \citep{passos}.
This phenomenological approach to introducing relaxation rates has
recently gotten additional motivation \citep{holder2019}.

\section{the Second order divergence in the DC third order response of gapped
graphene\label{sec:III}}

Having determined the analytical expression for the second order divergence
of the DC  third order conductivity, we can compare this result with
those that follow from a numerical calculation of the conductivity
in the velocity gauge. The material to be considered here is the gapped
graphene monolayer, described by a nearest neighbours tight-binding
model with parameters $\Delta=300\,$meV and $t=3\,$eV \citep{Ventura2019}.
For the analytical calculation, we consider an expansion of the tight-binding
Hamiltonian around the band minima, $\mathbf{k}=\mathbf{K}(\mathbf{K}')+\mathbf{q}$,
Fig.(\ref{fig:1}), which renders the usual Hamiltonian,
\begin{equation}
H_{\lambda}(\mathbf{q})=\left[\begin{array}{cc}
\Delta/2 & \ \hbar v_{F}(\lambda q_{x}-iq_{y})\\
\ \text{c.c.} & -\Delta/2
\end{array}\right]
\end{equation}
where $\lambda=\pm1$ for $\mathbf{K}=4\pi/3\sqrt{3}a_{0}\,\hat{k}_{x}$
and $\mathbf{K}'=-4\pi/3\sqrt{3}a_{0}\,\hat{k}_{x}$, respectively,
and $\hbar v_{F}=3ta_{0}/2$, for $a_{0}=1.42\text{Å}$, the distance
between two neighbouring atoms. Since this model has time reversal
symmetry, one has $\epsilon_{-\mathbf{k}s}=\epsilon_{\mathbf{k}s}$
and can choose the Berry connections such that, $\xi_{-\mathbf{k}ss'}^{\alpha_{1}}=\xi_{\mathbf{k}s's}^{\alpha_{1}}$.
One can then show that the only relevant portion of Eq.(\ref{eq:SYM})
is the one involving the symmetric product of Berry connections ---
$\xi_{\mathbf{k}ss'}^{\alpha_{1}}\xi_{\mathbf{k}s's}^{\alpha_{2}}+\xi_{\mathbf{k}ss'}^{\alpha_{2}}\xi_{\mathbf{k}s's}^{\alpha_{1}}$
--- so that the $\iota_{\beta\alpha_{1}\alpha_{2}\alpha_{3}}(\omega_{1})$
coefficient is necessarily real.

\begin{figure}[H]
\centering{}\includegraphics[width=0.96\columnwidth]{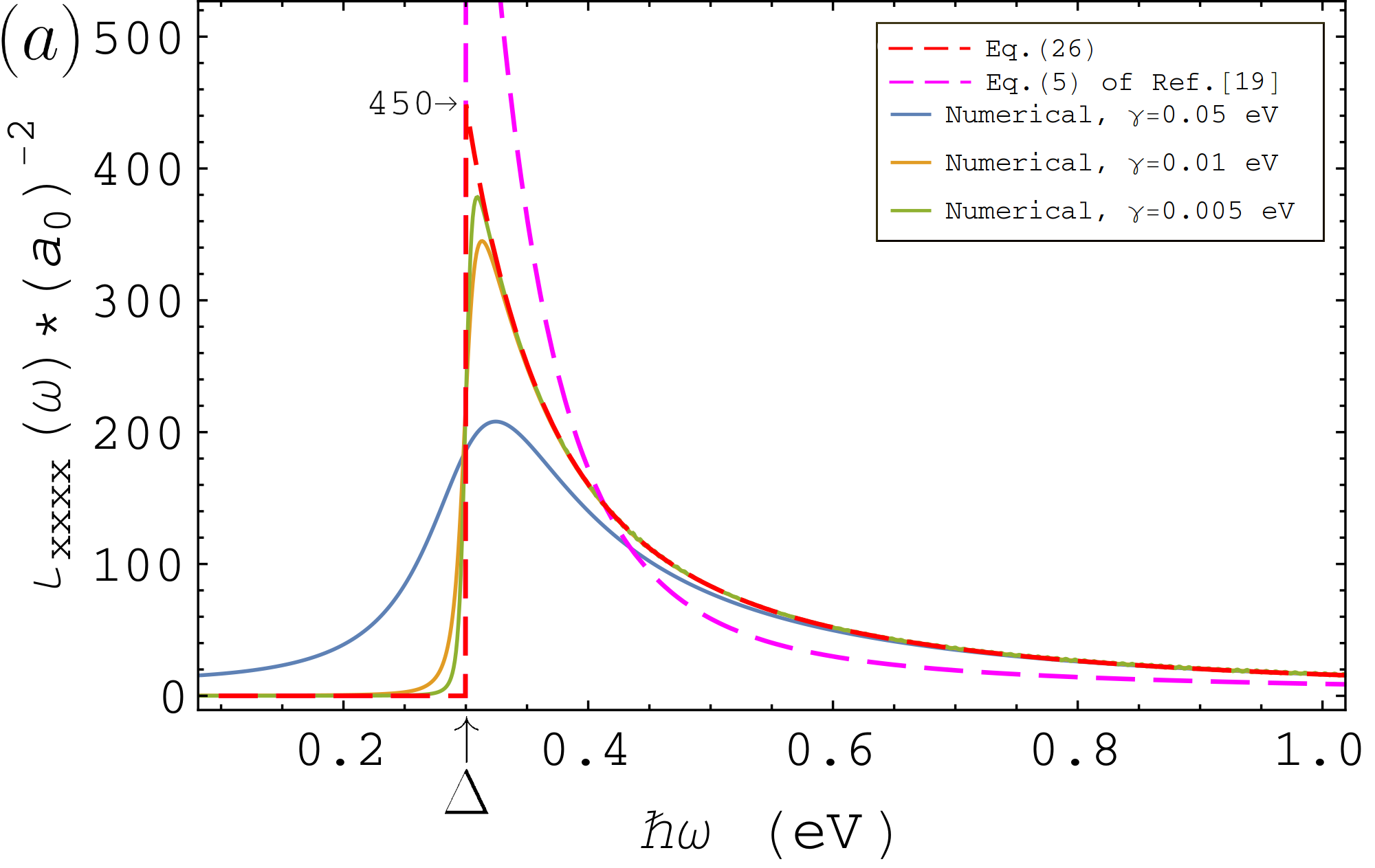}\medskip{}
\includegraphics[width=0.96\columnwidth]{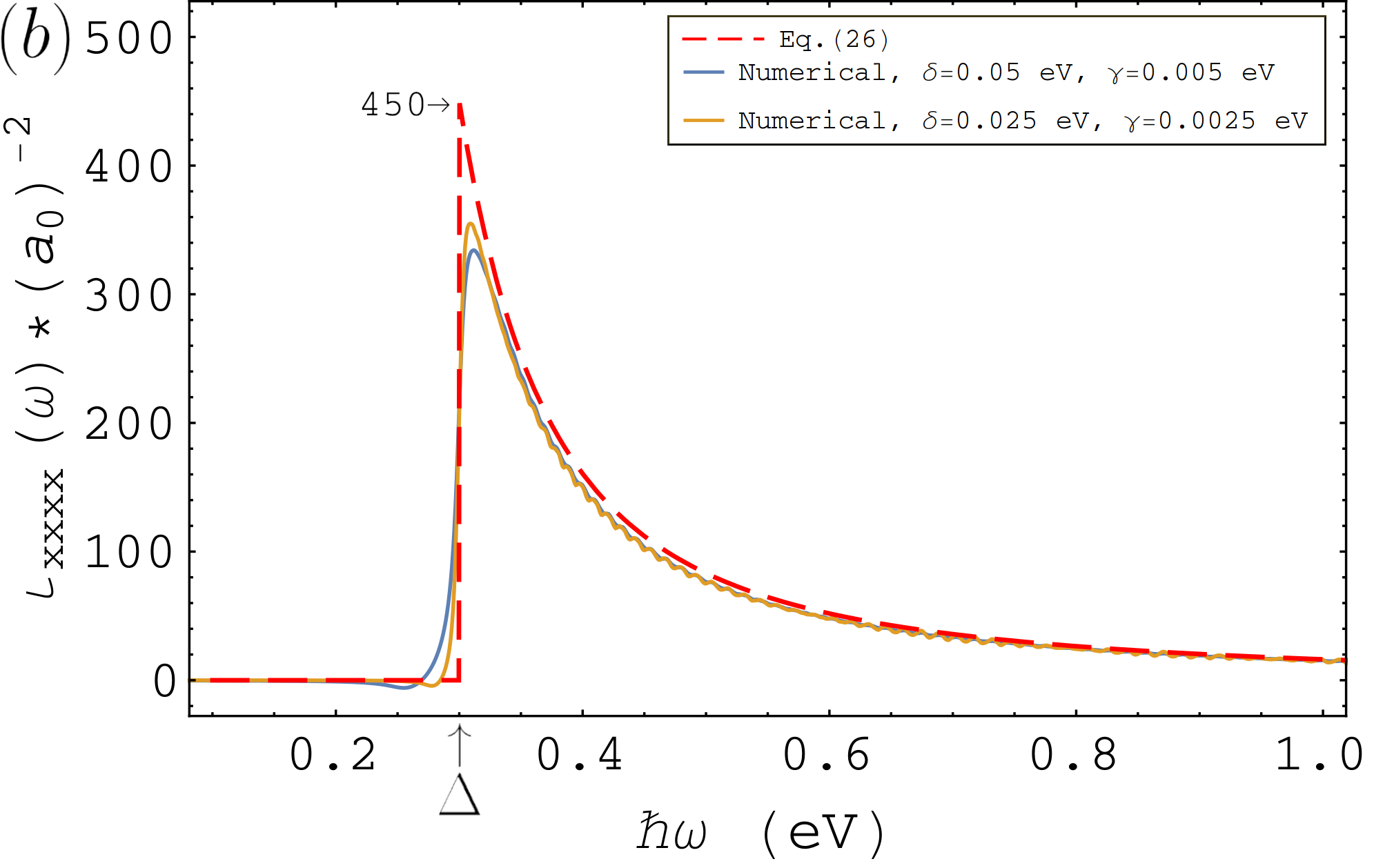}\medskip{}
\includegraphics[width=0.96\columnwidth]{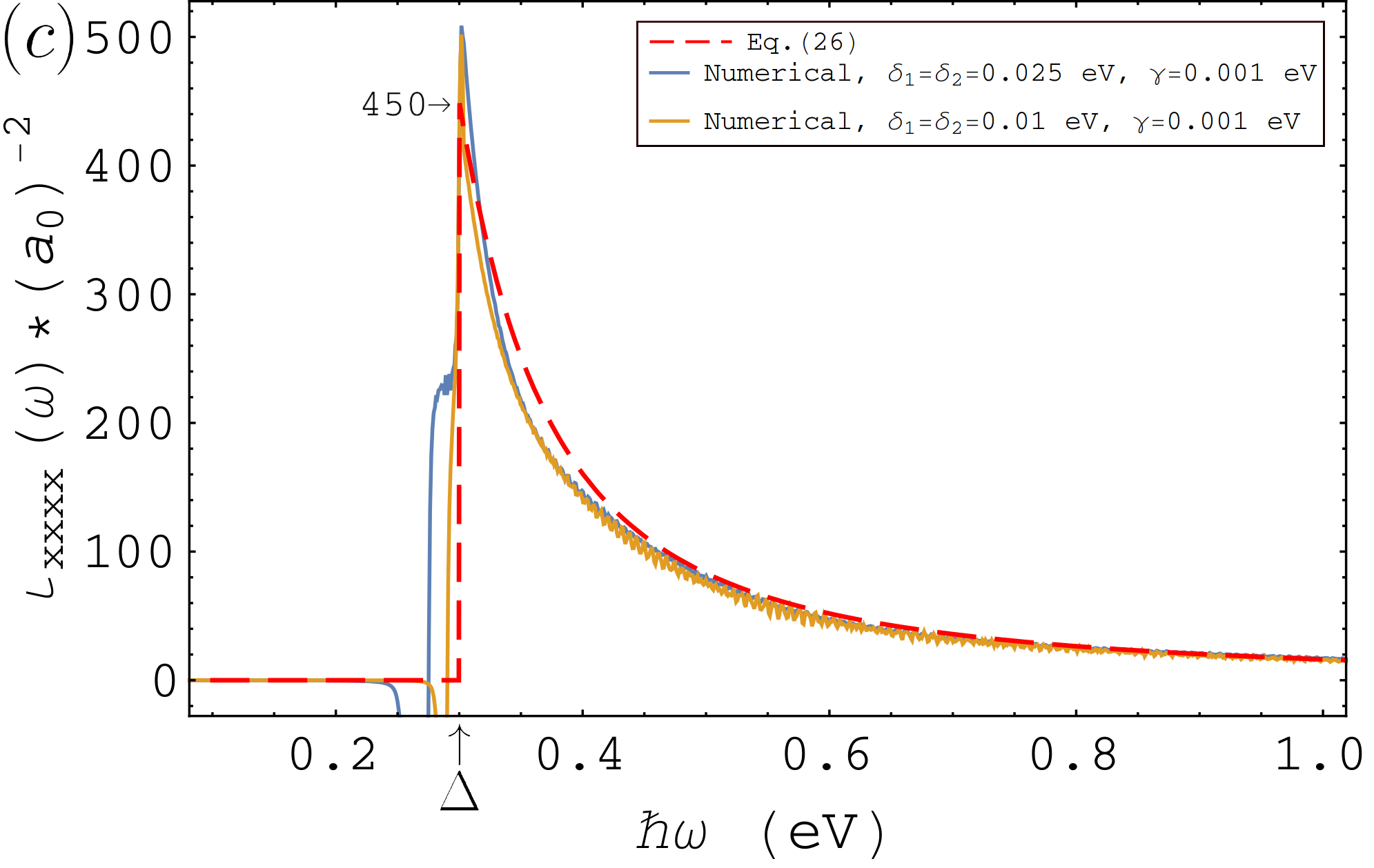}\caption{A comparison between the analytical result that follows from Eq.(\ref{eq:SYM})
--- represented by the red dashed curve --- and numerical results
from a velocity gauge calculation of the conductivity \citep{passos,Ventura2019}
--- represented by the full curves --- in gapped graphene, $\Delta=300\,$meV
and $t=3\,$eV, for the response along the zig-zag direction, $\beta=\alpha_{i}=x$,
$i=1,2,3$. The (a), (b) and (c) plots represent the second order
divergence in the jerk current and the dichromatic and trichromatic
field setup probes, respectively. In (a), we have also represented
the analytical solution from ref.\citep{fregoso2018} --- represented
by the magenta dashed curve. This, we note, does not match with our
numerical result. Note also that value of $\iota$ at the gap, given
by Eq.(\ref{eq:PRED}), reads as $\iota_{xxxx}(\Delta)a_{0}^{-2}=450$.
\label{fig:2}}
\end{figure}

The results for the three different probes to the divergence, Eqs.(\ref{eq:JERK_D})--(\ref{eq:DF_D}),
are presented in the three plots of Figure \ref{fig:2}. In the numerical
computation, the relaxation rate $\gamma$ is finite, as well as the
frequency offsets $\delta_{1}$ and $\delta_{2}$; $\delta_{1}$ for
the dichromatic field setup probe, plot (b), and both $\delta_{1}$
and $\delta_{2}$ for the trichromatic one, plot (c). So we compute
$\iota_{\beta\alpha_{1}\alpha_{2}\alpha_{3}}(\omega_{1})$ by expressing
it in terms of the conductivity, Eq.(\ref{eq:COND_EXP}),
\begin{equation}
\iota_{\beta\alpha_{1}\alpha_{2}\alpha_{3}}(\omega_{1})\approx-\frac{3\hbar^{3}}{e^{4}}\bar{\omega}_{123}\bar{\omega}_{12}\,\tilde{\sigma}_{\beta\alpha_{1}\alpha_{2}\alpha_{3}}(\omega_{1},\omega_{2},\omega_{3})
\end{equation}
and replacing $\bar{\omega}_{123},\ \bar{\omega}_{12}$ by the expressions
given in Eqs.(\ref{eq:JERK_D}) to (\ref{eq:DF_D}). There is good
agreement between the analytical results that follow from Eq.(\ref{eq:SYM}),
and the numerical results of a velocity gauge calculation of the \textit{full}
conductivity, for frequencies above the gap. This validates the analytical
expression that we have derived here. In addition, the results themselves
warrant two comments. First, we note that there are discrepancies
between the analytical and numerical results for frequencies below
the gap. These follow from the fact that the numerical calculation
carries contributions other than $\tilde{\Gamma}_{\beta\alpha_{1}\alpha_{2}\alpha_{3}}$
--- the terms represented by the ellipsis in Eq.(\ref{eq:COND_EXP})
--- which necessarily contribute to the response. That the differences
between results are more noticeable in plots (b) and (c) of Figure
\ref{fig:2} is due to the existence, in the full conductivity, of
resonances at frequencies $\omega\pm\delta$ and $\omega\pm\delta$,
$\omega+2\delta$, respectively. Secondly, the analytical calculation
gives us an expression for that $\iota_{xxxx}(\Delta)$ that reads
as, 
\begin{align}
\frac{1}{a_{0}^{2}}\iota_{xxxx}(\Delta)= & \ \frac{9t^{2}}{2\Delta^{2}}.\label{eq:PRED}
\end{align}
The second order divergence in gapped graphene should be more pronounced
when the band gap is smaller, which is consistent with the results
of ref.\citep{cheng2019} and similar to what was obtained for the
second order response \citep{Ventura2019}. 

Finally, we present an estimation of the amplitude of the jerk current,
$J_{\text{jerk}}^{x}$, along the zig-zag direction in gapped graphene.
For the values of the hopping parameter and band gap presented above
and for $\hbar\omega\sim\Delta$, $\tau=1/\gamma\sim100\,\text{fs}$,
$E_{\omega}^{x}=10^{7}\,\text{V/m}$ and $E_{0}^{x}=10^{6}\,\text{V/m}$,
one obtains $J_{\text{jerk}}^{x}\approx12\,\text{A}/\text{m}$, which
should be within experimental reach \citep{fregoso2018}. 

\section{Summary\label{sec:IV}}

The study of divergences in nonlinear optical response functions provides
us with the knowledge that some NLO responses can be made \textit{large}
simply by the choice of certain field setups, which is certainly relevant
from the standpoint of nonlinear optics. It can also provides us with
some valuable intuition concerning the physics that is associated
to these processes, as it has been done in \citep{Aversa1996,khurgin1995,fregoso2018}.
We have shown here that the leading order divergence in the third
order DC (or quasi-DC) response of a cold semiconductor --- first
identified in ref.\citep{fregoso2018} --- can be probed via three
different electric field setups, and is described by a single coefficient,
Eq.(\ref{eq:SYM}), that involves only one Dirac delta function, i.e.,
are localized contributions in the FBZ. The differences between the
results of this calculation and that of ref.\citep{fregoso2018} were
also addressed here. Finally, we compared, for the gapped graphene
monolayer, the results that follow from Eq.(\ref{eq:SYM}) with results
that follow from a numerical calculation of the conductivity in the
velocity gauge: these are in clear agreement with each other. 

The authors acknowledge financing of Funda\c{c}\~{a}o da Ci\^{e}ncia
e Tecnologia, of COMPETE 2020 program in FEDER component (European
Union), through projects POCI-01-0145-FEDER-028887 and UID/FIS/04650/2013.

\onecolumngrid

\appendix

\section{$s'\protect\ne s$ terms\label{sec:B}}

This appendix presents the intermediate steps of the derivation of
the $s'\ne s$ contribution that is considered in subsection \ref{subsec:Interband-contributions}.
It follows from Eq.(\ref{eq:COM_CD}) that the terms, $s'\ne s$,
Eq.(\ref{eq:INTERBAND}), can be written as the sum of two contributions,
\begin{multline}
(-1)\bigl[D_{\mathbf{k}}^{\alpha_{3}},v_{\mathbf{k}}^{\beta}\circ\frac{1}{\hbar\bar{\omega}_{123}+\Delta\epsilon_{\mathbf{k}}}\bigr]_{s's}\frac{1}{\hbar\bar{\omega}_{12}-\Delta\epsilon_{\mathbf{k}ss'}}\bigl[D_{\mathbf{k}}^{\alpha_{2}},\frac{(-i)}{\hbar\bar{\omega}_{1}-\Delta\epsilon_{\mathbf{k}}}\circ\bigl[\xi_{\mathbf{k}}^{\alpha_{1}},f_{\mathbf{k}}\bigr]\bigr]_{ss'}=\\
i\bigl[D_{\mathbf{k}}^{\alpha_{3}},v_{\mathbf{k}}^{\beta}\circ\frac{1}{\hbar\bar{\omega}_{123}+\Delta\epsilon_{\mathbf{k}}}\bigr]_{s's}\frac{1}{\hbar\bar{\omega}_{12}-\Delta\epsilon_{\mathbf{k}ss'}}\bigl(\nabla_{\mathbf{k}}^{\alpha_{2}}\frac{\xi_{\mathbf{k}ss'}^{\alpha_{1}}\Delta f_{\mathbf{k}s's}}{\hbar\bar{\omega}_{1}-\Delta\epsilon_{\mathbf{k}ss'}}\bigr)\\
+\bigl[D_{\mathbf{k}}^{\alpha_{3}},v_{\mathbf{k}}^{\beta}\circ\frac{1}{\hbar\bar{\omega}_{123}+\Delta\epsilon_{\mathbf{k}}}\bigr]_{s's}\frac{1}{\hbar\bar{\omega}_{12}-\Delta\epsilon_{\mathbf{k}ss'}}\bigl[\xi_{\mathbf{k}}^{\alpha_{2}},\frac{1}{\hbar\bar{\omega}_{1}-\Delta\epsilon_{\mathbf{k}}}\circ\bigl[\xi_{\mathbf{k}}^{\alpha_{1}},f_{\mathbf{k}}\bigr]\bigr]_{ss'}.\label{eq:B1}
\end{multline}
Let us take the first term on the RHS of Eq.(\ref{eq:B1}) and further
manipulate it. Since $\nabla_{\mathbf{k}}^{\alpha_{2}}f_{\mathbf{k}s}=0$
for a cold semiconductor, this term reads,
\begin{multline}
i\bigl[D_{\mathbf{k}}^{\alpha_{3}},v_{\mathbf{k}}^{\beta}\circ\frac{1}{\hbar\bar{\omega}_{123}+\Delta\epsilon_{\mathbf{k}}}\bigr]_{s's}\frac{\xi_{\mathbf{k}ss'}^{\alpha_{1}}\Delta f_{\mathbf{k}s's}}{\hbar\bar{\omega}_{12}-\Delta\epsilon_{\mathbf{k}ss'}}\left(\nabla_{\mathbf{k}}^{\alpha_{2}}\Delta\epsilon_{\mathbf{k}ss'}\right)\frac{1}{(\hbar\bar{\omega}_{1}-\Delta\epsilon_{\mathbf{k}ss'})^{2}}\\
+i\bigl[D_{\mathbf{k}}^{\alpha_{3}},v_{\mathbf{k}}^{\beta}\circ\frac{1}{\hbar\bar{\omega}_{123}+\Delta\epsilon_{\mathbf{k}}}\bigr]_{s's}\frac{\Delta f_{\mathbf{k}s's}}{\hbar\bar{\omega}_{12}-\Delta\epsilon_{\mathbf{k}ss'}}\bigl(\nabla_{\mathbf{k}}^{\alpha_{2}}\xi_{\mathbf{k}ss'}^{\alpha_{1}}\bigr)\frac{1}{\hbar\bar{\omega}_{1}-\Delta\epsilon_{\mathbf{k}ss'}}.\label{eq:B3}
\end{multline}
We will take a look at the first contribution Eq.(\ref{eq:B3}), as
it has been reported that terms like it --- which involve the product
of two denominators, one of them squared --- contribute to the second
order divergence of the third order DC conductivity \citep{fregoso2018,fregoso2019}.
It reads as,
\begin{equation}
i\bigl[D_{\mathbf{k}}^{\alpha_{3}},v_{\mathbf{k}}^{\beta}\circ\frac{1}{\hbar\bar{\omega}_{123}+\Delta\epsilon_{\mathbf{k}}}\bigr]_{s's}\xi_{\mathbf{k}ss'}^{\alpha_{1}}\Delta f_{\mathbf{k}s's}\bigl(\nabla_{\mathbf{k}}^{\alpha_{2}}\Delta\epsilon_{\mathbf{k}ss'})\frac{1}{\hbar\bar{\omega}_{12}-\Delta\epsilon_{\mathbf{k}ss'}}\frac{1}{(\hbar\bar{\omega}_{1}-\Delta\epsilon_{\mathbf{k}ss'})^{2}}.\label{eq:B4}
\end{equation}
We will show that these, in fact, do not to contribute.

\section{$s'=s$ terms\label{sec:A}}

This appendix presents the derivation of the $s'=s$ contribution
that is considered in subsection \ref{subsec:Intraband-contributions}.
It follows from Eq.(\ref{eq:COM_CD}) that Eq.(\ref{eq:INTRABAND})
can be expressed as,
\begin{multline}
(-1)\sum_{s}\bigl[D_{\mathbf{k}}^{\alpha_{3}},v_{\mathbf{k}}^{\beta}\circ\frac{1}{\hbar\bar{\omega}_{123}+\Delta\epsilon_{\mathbf{k}}}\bigr]_{ss}\frac{1}{\hbar\bar{\omega}_{12}-\Delta\epsilon_{\mathbf{k}ss}}\bigl[D_{\mathbf{k}}^{\alpha_{2}},\frac{(-i)}{\hbar\bar{\omega}_{1}+\Delta\epsilon_{\mathbf{k}}}\circ\bigl[\xi_{\mathbf{k}}^{\alpha_{1}},f_{\mathbf{k}}\bigr]\bigr]_{ss}=\\
\frac{(-1)}{\hbar\bar{\omega}_{12}}\sum_{s}\bigl[D_{\mathbf{k}}^{\alpha_{3}},v_{\mathbf{k}}^{\beta}\circ\frac{1}{\hbar\bar{\omega}_{123}+\Delta\epsilon_{\mathbf{k}}}\bigr]_{ss}\left(\frac{(-i)}{\hbar\bar{\omega}_{1}}\bigl(\nabla^{\alpha_{2}}\xi_{\mathbf{k}ss}^{\alpha_{1}}\Delta f_{\mathbf{k}ss}\bigr)+\bigl[\xi_{\mathbf{k}}^{\alpha_{2}},\frac{(-i)^{2}}{\hbar\bar{\omega}_{1}-\Delta\epsilon}\circ\bigl[\xi_{\mathbf{k}}^{\alpha_{1}},f_{\mathbf{k}}\bigr]\bigr]_{ss}\right)=\\
\frac{1}{\hbar\bar{\omega}_{12}}\sum_{s}\bigl[D_{\mathbf{k}}^{\alpha_{3}},v_{\mathbf{k}}^{\beta}\circ\frac{1}{\hbar\bar{\omega}_{123}+\Delta\epsilon_{\mathbf{k}}}\bigr]_{ss}\bigl[\xi_{\mathbf{k}}^{\alpha_{2}},\frac{1}{\hbar\bar{\omega}_{1}-\Delta\epsilon}\circ\bigl[\xi_{\mathbf{k}}^{\alpha_{1}},f_{\mathbf{k}}\bigr]\bigr]_{ss}.\label{eq:A22}
\end{multline}
In going from the second to the third line of Eq.(\ref{eq:A22}) we
used $\Delta f_{\mathbf{k}ss}:=f_{\mathbf{k}s}-f_{\mathbf{k}s}=0$.
What remains can then be expressed as,
\begin{align}
\frac{1}{\hbar\bar{\omega}_{12}}\sum_{s} & \left(\frac{1}{\hbar\bar{\omega}_{123}}\bigl(\nabla^{\alpha_{3}}v_{\mathbf{k}ss}^{\beta}\bigr)-i\bigl[\xi_{\mathbf{k}}^{\alpha_{3}},v^{\beta}\circ\frac{1}{\hbar\bar{\omega}_{123}+\Delta\epsilon}\bigr]_{ss}\right)\bigl[\xi_{\mathbf{k}}^{\alpha_{2}},\frac{1}{\hbar\bar{\omega}_{1}-\Delta\epsilon}\circ\bigl[\xi_{\mathbf{k}}^{\alpha_{1}},f_{\mathbf{k}}\bigr]\bigr]_{ss}.
\end{align}
It is now easy to identify the only term contributing to the second
order divergence in the DC response. Since, 
\begin{align}
\bigl[\xi_{\mathbf{k}}^{\alpha_{3}},v^{\beta}\circ\frac{1}{\hbar\bar{\omega}_{123}+\Delta\epsilon}\bigr]_{ss} & =\sum_{r}\left(\xi_{\mathbf{k}sr}^{\alpha_{3}}v_{rs}^{\beta}\frac{1}{\hbar\bar{\omega}_{123}+\Delta\epsilon_{rs}}-(s\leftrightarrow r)\right)=\sum_{r\left(\ne s\right)}\left(\xi_{\mathbf{k}sr}^{\alpha_{3}}v_{rs}^{\beta}\frac{1}{\hbar\bar{\omega}_{123}+\Delta\epsilon_{rs}}-(s\leftrightarrow r)\right),
\end{align}
does not carry a divergent factor. We are thus left with a single
contribution,
\begin{align}
\frac{1}{\hbar\bar{\omega}_{123}}\frac{1}{\hbar\bar{\omega}_{12}}\sum_{s} & \bigl(\nabla^{\alpha_{3}}v_{\mathbf{k}ss}^{\beta}\bigr)\bigl[\xi_{\mathbf{k}}^{\alpha_{2}},\frac{1}{\hbar\bar{\omega}_{1}-\Delta\epsilon}\circ\bigl[\xi_{\mathbf{k}}^{\alpha_{1}},f_{\mathbf{k}}\bigr]\bigr]_{ss},
\end{align}
that can be written as, 
\begin{equation}
-\frac{1}{\hbar\bar{\omega}_{123}}\frac{1}{\hbar\bar{\omega}_{12}}\frac{1}{\hbar}\sum_{r\ne s}\bigl(\nabla^{\beta}\nabla^{\alpha_{3}}\epsilon_{\mathbf{k}s}\bigr)\left(\frac{\xi_{\mathbf{k}sr}^{\alpha_{1}}\xi_{\mathbf{k}rs}^{\alpha_{2}}\Delta f_{\mathbf{k}rs}}{\hbar\bar{\omega}_{1}-\Delta\epsilon_{\mathbf{k}sr}}-(s\leftrightarrow r)\right).\label{eq:A3}
\end{equation}

\section{Symmetrizing the third order DC conductivity in the context of a
divergent response\label{sec:C}}

The relevant physical object in a conductivity description of the
response satisfies intrinsic permutation symmetry \citep{Boyd:2008}:
\begin{align}
\tilde{\sigma}^{\beta\alpha_{1}\alpha_{2}\alpha_{3}}(\omega_{1},\omega_{2},\omega_{3})= & \ \frac{1}{3!}\bigl[\sigma^{\beta\alpha_{1}\alpha_{2}\alpha_{3}}(\omega_{1},\omega_{2},\omega_{3})+\sigma^{\beta\alpha_{2}\alpha_{1}\alpha_{3}}(\omega_{2},\omega_{1},\omega_{3})+\sigma^{\beta\alpha_{1}\alpha_{3}\alpha_{2}}(\omega_{1},\omega_{3},\omega_{2})\nonumber \\
 & +\sigma^{\beta\alpha_{3}\alpha_{2}\alpha_{1}}(\omega_{3},\omega_{2},\omega_{1})+\sigma^{\beta\alpha_{3}\alpha_{1}\alpha_{2}}(\omega_{3},\omega_{1},\omega_{2})+\sigma^{\beta\alpha_{2}\alpha_{3}\alpha_{1}}(\omega_{2},\omega_{3},\omega_{1})\bigr]\label{eq:symmetrization}
\end{align}
As we are interested in singling out the terms that have second order
divergences, Eqs.(\ref{eq:JERK_D})--(\ref{eq:DF_D}) --- when both
$\omega_{123}$ and $\omega_{12}$ go to zero (or are much smaller
than $\omega$) --- we need only concern us with the first two terms
in Eq.(\ref{eq:symmetrization}); the remaining ones will only show
have a first order divergences in $\omega_{123}$. Symmetrizing Eq.(\ref{eq:UNISYM})
with respect to the frequencies and indexes $(\alpha_{1},\omega_{1})\leftrightarrow(\alpha_{2},\omega_{2})$,
one obtains for the integrand,

\begin{equation}
\sum_{s\ne s'}\frac{1}{6}\bigl(\nabla^{\beta}\nabla^{\alpha_{3}}\Delta\epsilon_{\mathbf{k}ss'}\bigr)\Delta f_{\mathbf{k}s's}\left(\xi_{\mathbf{k}ss'}^{\alpha_{1}}\xi_{\mathbf{k}s's}^{\alpha_{2}}\left[\frac{1}{\hbar\omega_{1}-\Delta\epsilon_{\mathbf{k}ss'}}-i\pi\delta(\hbar\omega_{1}-\Delta\epsilon_{\mathbf{k}ss'})\right]+\xi_{\mathbf{k}ss'}^{\alpha_{2}}\xi_{\mathbf{k}s's}^{\alpha_{1}}\left[\frac{1}{\hbar\omega_{2}-\Delta\epsilon_{\mathbf{k}ss'}}-i\pi\delta(\hbar\omega_{2}-\Delta\epsilon_{\mathbf{k}ss'})\right]\right).
\end{equation}
Note that the integration associated with the denominators is a principal
value one. Swaping $s'\leftrightarrow s$ in the terms with $\omega_{2}$
gives us,
\begin{equation}
\sum_{s\ne s'}\frac{1}{6}\bigl(\nabla^{\beta}\nabla^{\alpha_{3}}\Delta\epsilon_{\mathbf{k}ss'}\bigr)\Delta f_{\mathbf{k}s's}\xi_{\mathbf{k}ss'}^{\alpha_{1}}\xi_{\mathbf{k}s's}^{\alpha_{2}}\left[-i\pi\delta(\hbar\omega_{1}-\Delta\epsilon_{\mathbf{k}ss'})-i\pi\delta(-\hbar\omega_{2}-\Delta\epsilon_{\mathbf{k}ss'})+\frac{1}{\hbar\omega_{1}-\Delta\epsilon_{\mathbf{k}ss'}}+\frac{1}{\hbar\omega_{2}+\Delta\epsilon_{\mathbf{k}ss'}}\right].\label{eq:Sym_divergence}
\end{equation}
Now, let us consider the different cases in which we are interested.
For the jerk current, Eq.(\ref{eq:JERK_D}) --- $\omega_{1}=-\omega_{2}=\omega$,
$\omega_{3}=0$ --- we can see that the terms with the Delta functions
combine,
\begin{align}
-i\pi\delta(\hbar\omega-\Delta\epsilon_{\mathbf{k}ss'})-i\pi\delta(-\hbar(-\omega)-\Delta\epsilon_{\mathbf{k}ss'})= & -2i\pi\delta(\hbar\omega-\Delta\epsilon_{\mathbf{k}ss'}),
\end{align}
while those associated with principal value integrals cancel out,
\begin{align}
\frac{1}{\hbar\omega-\Delta\epsilon_{\mathbf{k}ss'}}+\frac{1}{-\hbar\omega+\Delta\epsilon_{\mathbf{k}ss'}}= & \ 0.
\end{align}
The same is true in the case of the dichromatic setup, Eq.(\ref{eq:MIXED_D})
--- $\omega_{1}=-\omega_{2}=\omega$ and $\omega_{3}=\delta$ ---
as only $\omega_{3}$ changes. For the trichromatic setup, Eq.(\ref{eq:DF_D})
--- $\omega_{1}=\omega+\delta_{1}$, $\omega_{2}=-\omega$ and $\omega_{3}=\delta_{2}$
--- the resonances do contribute with additional terms, but, since
we already isolated the terms that diverge as $\delta_{1}\to0$ we
can replace the term in square brackets in Eq.(\ref{eq:Sym_divergence})
by its value at $\delta_{1}=0$,
\begin{align}
-i\pi\delta(\hbar(\omega+\delta_{1})-\Delta\epsilon_{\mathbf{k}ss'})-i\pi\delta(-\hbar(-\omega)-\Delta\epsilon_{\mathbf{k}ss'})+\frac{1}{\hbar(\omega+\delta_{1})-\Delta\epsilon_{\mathbf{k}ss'}}+\frac{1}{\hbar(-\omega)+\Delta\epsilon_{\mathbf{k}ss'}}\to & -2i\pi\delta(\hbar\omega-\Delta\epsilon_{\mathbf{k}ss'}).
\end{align}
As before, the only relevant terms are those associated with a Dirac
delta function in the frequency $\omega$. We can thus write the symmetrized
contribution to the second order divergence of the third order DC
conductivity, $\tilde{\Gamma}_{\beta\alpha_{1}\alpha_{2}\alpha_{3}}(\omega_{1},\omega_{2},\omega_{3})$,
as,
\begin{align}
\tilde{\Gamma}_{\beta\alpha_{1}\alpha_{2}\alpha_{3}}(\omega_{1},\omega_{2},\omega_{3})= & -\frac{\pi e^{4}}{3\hbar^{3}}\frac{1}{\bar{\omega}_{123}}\frac{1}{\bar{\omega}_{12}}\int\frac{d^{d}\mathbf{k}}{(2\pi)^{d}}\sum_{s\ne s'}\bigl(\nabla^{\beta}\nabla^{\alpha_{3}}\Delta\epsilon_{\mathbf{k}ss'}\bigr)\xi_{\mathbf{k}ss'}^{\alpha_{1}}\xi_{\mathbf{k}s's}^{\alpha_{2}}\Delta f_{\mathbf{k}s's}\,\delta(\hbar\omega-\Delta\epsilon_{\mathbf{k}ss'}).
\end{align}

\twocolumngrid

%\bibliographystyle{apsrev4-1}
%\bibliography{NLO}

%

\end{document}